\documentclass[nofootinbib,twocolumn,showpacs,preprintnumbers,floatfix,amsmath,amssymb,floatfix]{revtex4}

\usepackage{graphicx} 
\usepackage{dcolumn} 
\usepackage{bm} 
\def\pslash{p\!\!\!\slash }
\def\Zslash{Z\!\!\!\!\slash }

\begin{document}

\preprint{}
\title{Spin Content of $\Lambda$ in QCD Sum Rules}
 \author{G\"{u}ray Erkol}
 \email{guray.erkol@ozyegin.edu.tr}
 \affiliation{Laboratory for Fundamental Research, Ozyegin University, Kusbakisi Cad. No:2 Altunizade Uskudar, Istanbul 34662 Turkey}
 \author{Makoto Oka}%
 \email{oka@th.phys.titech.ac.jp}
 \affiliation{Department of Physics, H27, Tokyo Institute of Technology, Meguro, Tokyo 152-8551 Japan}
\begin{abstract} 
	We calculate the isoscalar axial-vector coupling constants of the $\Lambda$ hyperon using the method of QCD sum rules. A determination of these coupling constants reveals the individual contributions of the $u$, $d$ and the $s$ quarks to the spin content of $\Lambda$. Our results for the light-quark contributions are in agreement with those from experiment assuming flavor SU(3). We also find that the flavor-SU(3)-breaking effects are small and the contributions from the $u$ and the $d$ quarks to the $\Lambda$ polarization are negatively polarized as in the flavor-SU(3) limit. 
\end{abstract}
\pacs{12.38.Lg, 14.20.Jn, 13.88+e} \keywords{}

\maketitle
\section{Introduction}
According to data from polarized lepton-nucleon deep inelastic scattering~(DIS) which was initially reported by the EMC Collaboration~\cite{Ashman:1987hv} and confirmed subsequently by several other experiments~\cite{Adeva:1998vv, Abe:1998wq, Airapetian:2007mh}, only a small fraction of the nucleon spin is carried by the valence quarks. This observation has triggered many research activities and puzzles in understanding the spin content of the nucleon (see \emph{e.g.} Refs.~\cite{Filippone:2001ux,Bass:2004xa} for a review). In this respect, it has been realized that the polarized $\Lambda$, having all spin carried by the $s$-quark while the $u$-$d$ quark pair is coupled to $S=0$, $I=0$, provides a special example in the naive quark model. Contrary to the naive expectations, the interpretation of the experimental data together with the SU(3)$_F$ symmetry is that $\sim 60\%$ of the $\Lambda$ spin is carried by the $s$ (and $\bar{s}$) quark, while $\sim -40\%$ originates from $u$ (and $\bar{u}$) and $d$ (and $\bar{d}$) quarks~\cite{Burkardt:1993zh}. One important aspect of this interpretation is that the $u$-$d$ quark pair has negative polarization. Actually, this result is supported by some model-dependent approaches as well~(see Appendix~\ref{appA}). An investigation of this interesting issue has the potential to shed light on the ``spin crisis'' and therefore attracted considerable attention~\cite{Burkardt:1993zh,Jaffe:1996wp,Boros:1998kc,Ashery:1999am,Yang:2001pr,Gockeler:2002uh}. Experimentally, the polarization of $\Lambda$ is of special interest because it can be easily measured from the nonleptonic decay $\Lambda\rightarrow p\,\pi$~\cite{Buskulic:1996vb, Ackerstaff:1997nh, Airapetian:1999sh, Adams:1999px,Airapetian:2006ee}.

One open question in this framework is how sensitive the $\Lambda$ spin structure to SU(3)$_F$ breaking effects is. While it is claimed that the SU(3)$_F$ breaking may lead to a change in the sign of the $u$- and the $d$-quark polarizations in $\Lambda$~\cite{Yang:2001pr}, lattice~\cite{Gockeler:2002uh} and some model-dependent works~\cite{Ashery:1999am} find that $\Lambda$ is insensitive to SU(3)$_F$ breaking. The isoscalar ($g_A^q$ and $g_A^s$), octet ($g_8^q$) and singlet ($g_0^q$) axial-vector coupling constants of $\Lambda$ (shown generically as $g_A\equiv g_A(q^2=0)$ throughout the text) can be expressed in terms of the fractional contributions of the quark flavors, $\Delta q$, to the $\Lambda$ spin content as
\begin{align}
	g_A^q &= \Delta u + \Delta d,&\quad g_A^s &= \Delta s,\\
	g_A^8 &= \Delta u + \Delta d -2 \Delta s,&\quad g_A^0 &= \Delta u + \Delta d + \Delta s,\nonumber
\end{align}
in the SU(3)$_F$ limit. Therefore, a determination of these coupling constants reveals the spin content of $\Lambda$. 

Our primary aim in this paper is to calculate the isoscalar coupling constants $g_A^q$ and $g_A^s$. For this purpose we use the method of QCD sum rules (QCDSR)~\cite{Shifman:1978bx,Shifman:1978by,Reinders:1984sr,Ioffe:1983ju}. Note that this is reminiscent of the works in Refs.~\cite{Belyaev:1984vv,Chiu:1985ey,Gupta:1988nba,Henley:1992vq,Lee:1996dc,Ioffe:1998sa,Pasupathy:2003tf}, where the axial-vector coupling constants of the nucleon have been calculated. Using this method, we also extract the contributions of the $u$ and the $d$ quarks to $\Lambda$ spin content in the SU(3)$_F$-breaking case in order to see how sensitive the results are to symmetry-breaking effects.

We have organized our paper as follows: In section~\ref{form} we derive the QCD sum rules for $g_A$ and give numerical analysis and the results in section~\ref{res}. Finally we conclude in section~\ref{conc}.

\section{Formulation}\label{form}
We start with the correlation function of two $\Lambda$ interpolating fields in the presence of an external constant axial-vector field $Z_\mu$, defined by 
\begin{align}
	\begin{split}\label{tpc}
		&i\int d^4 x~ e^{i p\cdot x}\, \left \langle 0\left \lvert{\cal 
		T}[\eta_\Lambda(x)\overline{\eta}_\Lambda(0)]\right\rvert 0\right\rangle_Z=\\
		&\qquad\Pi(p) + Z\cdot\Pi^Z (p)+ O(Z^2).
	\end{split}	
\end{align}
This correlation function is computed by adding the term
\begin{equation}
	\Delta{\cal L}=-\sum_q g_q\, \overline{q}\, \Zslash\, \gamma_5\, q,
\end{equation}
to the usual QCD Lagrangian, where $g_q$ is the coupling of the quark field to the external field and we use the notation $\Zslash=Z^\mu \gamma_\mu$. 

The most general $\Lambda$ interpolating field is defined as a mixture of two independent local operators via the mixing parameter $t$:
	\begin{align} 	
		\begin{split}\label{intfi}
		\eta_{\Lambda} =& -\sqrt{\frac{2}{3}} \, \epsilon_{abc}\Big\{2\left[u_a^T C\gamma_5 d_b\right]s_c + \left[u_a^T C\gamma_5 s_b\right]d_c  \\
		&\quad -\left[d_a^T C\gamma_5 s_b\right]u_c +t \left(2 \left[u_a^T C d_b\right]\gamma_5 s_c \right.\\
		&\quad \left.+ \left[u_a^T C s_b\right]\gamma_5 d_c - \left[d_a^T C s_b\right]\gamma_5 u_c\right)\Big\},
		\end{split}
	\end{align}%
where $a,b,c$ are the color indices, $T$ denotes transposition and $C=i\gamma^2\gamma^0$. The choice $t=-1$ gives the Ioffe's current, which is often used in QCDSR calculations. In our numerical analysis, we take $t=-1.2$ which produces the optimal interpolating field~\cite{Leinweber:1995fn}. In Eq.~\eqref{tpc}, $\Pi(p)$ is the correlation function when the external field is absent and corresponds to the function that is used to determine the $\Lambda$ mass, while $\Pi^Z (p)$ represents the linear response of the correlator to a small external axial-vector field $Z_\mu$. In the presence of an external axial-vector field, Lorentz invariance of the vacuum is broken and new vacuum condensates appear as
\begin{align}
	&\langle\bar{q} \gamma_\mu\gamma_5 q \rangle_Z = g_q\,  Z_\mu\, \chi\, \langle \bar{q} q \rangle,\\
	&\langle \bar{q} g_c \tilde{G}_{\mu\nu} \gamma^\nu q \rangle_Z = g_q\, Z_\mu\, \kappa\, \langle \bar{q}q\rangle, \quad \tilde{G}_{\mu\nu}=\frac{1}{2}\epsilon_{\mu\nu\alpha\beta}\,G^{\alpha\beta}, 
\end{align}
which are defined in terms of the susceptibilities $\chi$ and $\kappa$ with the QCD coupling-constant squared $g_c^2=4\pi\alpha_s$.

We can bring the correlation function in Eq.~\eqref{tpc} into the form
	\begin{align}
		\Pi(p)&=\Pi_1(p^2) + \Pi_2(p^2)\,\pslash,\\	
	\begin{split}
		Z\cdot\Pi^Z (p)&=\Pi^Z_1(p^2)\,iZ_\mu\,\sigma^{\mu\nu}\, p_\nu\gamma_5  \\
		&\quad + \Pi^Z_2(p^2)\, Z\cdot p\, \pslash\, \gamma_5+ \Pi^Z_3(p^2)\, \Zslash\,\gamma_5.
	\end{split}
	\end{align}%
The Operator Product Expansion (OPE) sides of the sum rules are obtained by inserting the interpolating field \eqref{intfi} into the correlation function \eqref{tpc} and evaluating the time-ordered contractions of quark fields, which include the quark propagators~\cite{Belyaev:1984vv,Chiu:1985ey,Henley:1992vq}. On the OPE side, we include the terms up to dimension 8. The perturbative-correction terms in order of $\alpha_s$ may become important especially at lower Borel-mass region but they are expected to give smaller contribution to the sum rule we choose to work with (see Section~\ref{res})~\cite{Leinweber:1995fn}. Therefore these correction terms are neglected in this work. The phenomenological side is obtained via a dispersion relation, which is written in terms of hadron degrees of freedom. Finally, the QCD sum rules are constructed by matching the OPE sides with the phenomenological sides and applying the Borel transformation. The details of this procedure can be found in the extensive literature on QCDSR. Omitting the details, here we give the final forms of the sum rules for $g_A$ as obtained at three different Lorentz-Dirac structures:
{\allowdisplaybreaks
\begin{align}\label{spinSR}
	\begin{split}
	&iZ_\mu\,\sigma^{\mu\nu}\, p_\nu\gamma_5:\\
	&\Big[C_1\, a_q\, M^4\, E_0\,L^{2/9} + C_2\, m_s\, \chi\, a_q\, M^4\, E_0\,L^{-8/9}  \\
	&\qquad + C_3\, \chi\, a_q^2\, M^2 + C_4\, \kappa\, a_q^2 \,L^{-32/81} \\
	&\qquad + C_5\, \chi\, m_0^2\, a_q^2\,L^{-14/27} + C_6\,a_q\,b + C_7\, m_s\, a_q^2\,\Big]\\
	&\qquad \times \frac{e^{m_\Lambda^2/M^2}}{\tilde{\lambda}_\Lambda^2\,m_\Lambda}\\
	&\quad=(g_A + A\, M^2),
	\end{split}\\
	\begin{split}\label{spinSR2}
	&Z\cdot p\, \pslash\, \gamma_5:\\
	&\Big[C_1^\prime\, M^6\, E_1\,L^{-4/9} + C_2^\prime\, \chi\, a_q\, M^4\, E_0\,L^{-4/9} \\
	&\qquad+ C_3^\prime\, \kappa\, a_q\,M^2 \,L^{-68/81}+ C_4^\prime\, b\,M^2 \,L^{-4/9}\\
	&\qquad + C_5^\prime\, m_s\, a_q\, M^2 \,L^{-4/9} + C_6^\prime\,a_q^2\,L^{4/9}  \\
	&\qquad + C_7^\prime\, \chi\,a_q\,b\,L^{-4/9} + C_8^\prime\, m_s\,\chi\, a_q^2\,L^{-4/9}\\
	&\qquad + C_9^\prime\,m_s\,\kappa\,\frac{a_q^2}{M^2}\,L^{-68/81} \Big]\frac{e^{m_\Lambda^2/M^2}}{\tilde{\lambda}_\Lambda^2}\\
	&\quad=(g_A + A^\prime\, M^2),
	\end{split}\\
	\begin{split}\label{spinSR3}
	&\Zslash\,\gamma_5:\\
	&\Big[\mathring{C}_1\, M^8\, E_2\,L^{-4/9} + \mathring{C}_2\, \chi\, a_q\, M^6\, E_1\,L^{-4/9} \\
	&\qquad+ \mathring{C}_3\, \kappa\, a_q\,M^4\,E_0\,L^{-68/81} + \mathring{C}_4\, b\,M^4\,E_0 \,L^{-4/9}\\
	 &\qquad +\mathring{C}_5\, m_s\, a_q\, M^4\, E_0 \,L^{-4/9}+ \mathring{C}_6\,a_q^2\,M^2 \,L^{4/9}  \\
	&\qquad + \mathring{C}_7\,\chi\,a_q\,b\,M^2\,L^{-4/9} + \mathring{C}_8\, m_s\,\chi\, a_q^2\,M^2\,L^{-4/9}\\
	&\qquad +\mathring{C}_9\,m_s\,\kappa\,a_q^2\,L^{-68/81}\Big] \frac{e^{m_\Lambda^2/M^2}}{\tilde{\lambda}_\Lambda^2\, (M^2-2 m_\Lambda^2)}\\
	&\quad=(g_A + \mathring{A}\, M^2),
\end{split}
\end{align}%
}
where we have defined
{\allowdisplaybreaks
	\begin{align}
	C_1&=-\frac{(t-1)}{18}\Big[(g_u+g_d)(5t+1)+g_s(t+3)f\Big],\notag\\
	C_2&=-\frac{(t-1)^2}{12}(g_u+g_d),\notag\\
	C_3&=-\frac{(t-1)}{18}\Big[(g_u+g_d)[(f-5)t-(f+1)]\notag\\
	&\qquad +2g_s(t+5)f\Big],\notag\\
	C_4&=\frac{(t-1)}{324}\Big[(g_u+g_d)[(13f-47)t+(5-13f)]\notag\\
	&\qquad +2g_s(47-5t)f\Big],\label{coefmass}\\
	C_5&=\frac{(t-1)}{144}\Big[(g_u+g_d)[-(3f+7)t-(5f+3)]\notag\\
	&\qquad +2g_s(t+5)f\Big],\notag\\
	C_6&=-\frac{(t-1)}{1296}\Big[(g_u+g_d)(11t+7)+g_s(35t+37)f\Big],\notag\\	
	C_7&=\frac{(t-1)}{54}(g_u+g_d)\Big[2(t-1)+(5t+1)f\Big],\notag
	\end{align} 
\begin{align}
C^\prime_1 &= \frac{1}{24}\Big[ (5 t^2 + 26 t + 5) (g_u + g_d) + (11 t^2 + 14 t + 11) g_s\Big],\notag\\
C^\prime_2 &= \frac{1}{18}\Big[-2 (t^2 + 7 t + 1) (g_u + g_d) + (t - 1)^2 f g_s\Big],\notag\\
C^\prime_3 &= -\frac{1}{108}\Big[(29 t^2 + 50 t + 29) (g_u + g_d)\notag\\
&\qquad + 2 (13 t^2 + 10 t + 13) f g_s\Big],\notag\\
C^\prime_4 &= \frac{1}{96}\Big[(t - 1)^2 (g_u + g_d) + (13 t^2 + 10 t + 13) g_s\Big],\label{coefmass2}\\
C^\prime_5 &= -\frac{1}{36}\Big[(5 t + 1) [4 (t - 1) + 3 (t + 5) f] (g_u + g_d)\Big],\notag\\
C^\prime_6 &= -\frac{(t-1)}{27} \Big[(10 f t + t + 2 f - 1) (g_u + g_d) \notag\\
&\qquad + (2 f t + 33 t + 10 f + 39) g_s\Big],\notag\\
C^\prime_7 &= \frac{1}{432}\Big[(7 t^2 + 4 t + 7) (g_u + g_d) + (t - 1)^2 f g_s\Big],\notag\\
C^\prime_8 &= \frac{1}{9}\Big[(t^2 + 7 t + 1) f (g_u + g_d)\Big],\notag\\
C^\prime_9 &= \frac{1}{243}\Big[[(7 f + 10) t^2 + (13 f - 8) t + 7 f - 2](g_u + g_d)\Big] ,\notag
	\end{align} 
	\begin{align}
\mathring{C}_1 &= \frac{1}{24}\Big[ (5 t^2 + 26 t + 5) (g_u + g_d) + (11 t^2 + 14 t + 11) g_s\Big],\notag\\
\mathring{C}_2 &= \frac{1}{18}\Big[ - (5 t^2 + 8 t + 5) (g_u+g_d) \notag\\
&\qquad- 2  (19 t^2 + 16 t + 19) f g_s\Big],\notag\\
\mathring{C}_3 &= \frac{1}{36}\Big[ [ (25 t^2 - 14 t + 25] (g_u+g_d)\notag\\
&\qquad + 2 ( 41 t^2 + 26t + 41) f g_s\Big],\notag\\
\mathring{C}_4 &= \frac{1}{96}\Big[(t - 1)^2 (g_u + g_d) + (13 t^2 + 10 t + 13) g_s\Big],\notag\\
\mathring{C}_5 &= -\frac{(5 t+1)}{12}\Big[ [5 f+t (f+4)-4] (g_u+g_d)\Big],\notag\\
\mathring{C}_6 &= -\frac{(t - 1)}{27}\Big[ [(4 f + t (20 f - 1) + 1] (g_u+g_d)\notag\\
&\qquad + [t (33 - 2 f) - 10 f + 39] g_s\Big],\label{coefmass3}\\
\mathring{C}_7 &= -\frac{1}{216}\Big[ (5 t^2 - t + 5] (g_u+g_d)\notag \\
&\qquad+ 2 (10 t^2 + 7 t + 10] f g_s\Big],\notag\\
\mathring{C}_8 &= \frac{1}{18}\Big[ [(3 f+10) t^2+ 4 (3 f-2) t + 3 f-2](g_u+g_d)\Big],\notag\\
\mathring{C}_9 &= \frac{1}{162}\Big[[(7 f + 10) t^2 + (13 f - 8) t + 7 f - 2](g_u + g_d)\Big].\notag
\end{align}%
}%
Here $M$ is the Borel mass and the overlap amplitude is defined via $\langle 0 \lvert \eta_\Lambda \rvert \Lambda(p) \rangle= \lambda_\Lambda \upsilon(p)$ [$\upsilon(p)$ is the Dirac spinor for $\Lambda$ with momentum $p$] with $\tilde{\lambda}_\Lambda^2=32 \pi^4 \lambda_\Lambda^2$. We have also defined the quark condensate $a_q=-(2\pi)^2\langle\overline{q}q\rangle$, and the quark-gluon--mixed condensate $\langle\overline{q}g_c {\bm \sigma} \cdot {\bf G} q\rangle=m_0^2 \langle\overline{q}q\rangle$. The flavor-symmetry breaking is accounted for by the factor $f=\langle\overline{s}s\rangle/\langle\overline{q}q\rangle$. The continuum contributions are included via the factors 
	\begin{equation}
		E_n\equiv 1- \left(1+x+...+ \frac{x^n}{n!}\right) e^{-x},
	\end{equation}%
with $x=w^2/M^2$, where $w$ is the continuum threshold. The corrections that come from the anomalous dimensions of various operators are included with the factors $L=\log(M^2/\Lambda_{QCD}^2)/\log(\mu^2/\Lambda_{QCD}^2)$, where $\mu$ is the renormalization scale and $\Lambda_{QCD}$ is the QCD scale parameter.

\section{Results}\label{res}
In principle one can use any of the three sum rules to calculate $g_A$, however, not all of them work equally well due to continuum effects and insufficient OPE convergence. For the calculation of the nucleon axial-vector coupling constants, the sum rule at the structure $Z\cdot p\, \pslash\, \gamma_5$ has been favored over the others in the literature. On the other hand, it has been found in Ref.~\cite{Lee:1996dc} that this sum rule fails to have a valid Borel region whereas the sum rule at the structure $iZ_\mu\,\sigma^{\mu\nu}\, p_\nu\gamma_5$ satisfies OPE convergence and pole dominance, therefore has a valid Borel window. 

The valid Borel regions are determined so that the highest-dimensional operator contributes no more than about 10$\%$ to the OPE side, which gives the lower limit and ensures OPE convergence. The upper limit is determined using a restrictive criterion such that the continuum-plus-excited-state contributions are less than about $30\%$ of the phenomenological side, which is imposed so as to warrant the pole dominance. Using these criteria we have similarly found that the sum rule in \eqref{spinSR} has a valid Borel window while those in \eqref{spinSR2} and \eqref{spinSR3} are seriously contaminated by continuum contributions. Therefore we choose to work with the sum rule at the structure $iZ_\mu\,\sigma^{\mu\nu}\, p_\nu\gamma_5$ in \eqref{spinSR}. In order to obtain the corresponding sum rules for $g^q_A$ ($g^s_A$) we set $g_u = g_d=1$~($g_u = g_d=0$) and $g_s=0$~($g_s=1$).

We determine the uncertainties in the extracted parameters via the Monte Carlo--based analysis introduced in Ref.~\cite{Leinweber:1995fn}. In this analysis, randomly selected, Gaussianly distributed sets are generated from the uncertainties in the QCD input parameters. Here we use $a_q=0.52 \pm 0.05$~GeV$^3$, $b\equiv\left\langle g_c^2 G^2\right\rangle=1.2 \pm 0.6$~GeV$^4$, $m_0^2=0.72 \pm 0.08$~GeV$^2$, and $\Lambda_{QCD}=0.15 \pm 0.04$~GeV. The flavor-symmetry breaking parameter and the mass of the $s$-quark are taken as $f\equiv \langle \overline{s}s \rangle/\langle \overline{u}u \rangle= 1$ and $m_s=0$, respectively, in the SU(3)$_F$ limit. In the SU(3)$_F$-broken case, we take these parameter values as $f= 0.83 \pm 0.05$ and $m_s=0.11 \pm 0.02$~GeV. The values of the vacuum susceptibilities have been estimated in Refs.~\cite{Belyaev:1984vv,Pasupathy:1986pw,Pasupathy:2003tf,Ioffe:2005ym}. We consider the values $\chi a_q=0.60$~GeV$^2$ and $\kappa a_q=0.05$~GeV$^4$ for $g_A^q$ and $g_A^s$. The continuum threshold is taken as $w=1.5$~GeV in the SU(3)$_F$ limit and as $w=1.7$~GeV in the SU(3)$_F$-breaking case.

\begin{figure}
	[th] 
	\caption{The left-hand (solid red curves) and the fitted right-hand (dotted lines) sides of the sum rules \eqref{spinSR} for $g_A^q$ and $g_A^s$ in their valid Borel regions. The bands show the errors as obtained from the Monte Carlo-based analysis and the diamonds mark the values based on expectations from experimental results assuming SU(3)$_F$.}
	\includegraphics[scale=0.50]{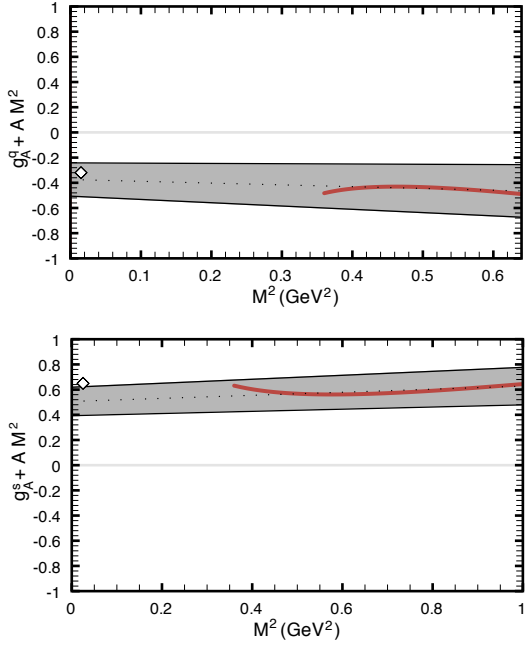}  \label{axials}
\end{figure}

For normalization of the sum rule \eqref{spinSR}, we use the chiral-odd mass sum rule, which is obtained using the invariant function $\Pi_1(p^2)$ as follows: 
\begin{align}\label{massSR}
	\begin{split}
	&\tilde{C}_1\, a_q\, M^4\, E_1 + \tilde{C}_2\, m_0^2\, a_q\, M^2\, E_0 + \tilde{C}_3\, a_q\, b \\
	&\qquad + \tilde{C}_4\, m_s\, M^6\, E_2 + \tilde{C}_5\, m_s\, a_q^2 + \tilde{C}_6\, m_s\, b\, M^2\,E_0\\
	&\quad=\frac{\tilde{\lambda}_\Lambda^2}{2}\,m_\Lambda\,e^{-m_\Lambda^2/M^2},
\end{split}
\end{align}%
where
{\allowdisplaybreaks
\begin{align}
	\begin{split}
	&\hat{C}_1=-\frac{(t-1)}{12}\Big[11f+10+(13f+2)t\Big],\\
	&\hat{C}_2=\frac{(t-1)}{24}\Big[7f+11+(11f+7)t\Big],\\
	&\hat{C}_3=-\frac{(t-1)}{288}\Big[13f+2+(11f+10)t\Big],\\
	&\hat{C}_4=-\frac{(t-1)}{12}(13t+11),\\
	&\hat{C}_5=\frac{1}{18}\Big[3(5t^2+2t+5)+(t-1)(t+5)f\Big],\\
	&\hat{C}_6=-\frac{(t-1)}{96}(11t+13).\\
\end{split}
\end{align}%
}%
Note that the chiral-odd mass sum rule has been found to be more reliable than the chiral-even one~\cite{Leinweber:1995fn}.

We first concentrate on the sum rules for the isoscalar coupling constants $g_A^q$ and $g_A^s$ in the SU(3)$_F$ limit. The Monte Carlo-based analyses of the sum rules are performed by first fitting the mass sum rule~\eqref{massSR} to simultaneously obtain $m_\Lambda$ and $\tilde{\lambda}_\Lambda$, and the obtained value for the overlap amplitude is used in the sum rules of $g_A$ for each corresponding parameter set. In Fig.~\ref{axials}, we plot the left-hand and the fitted right-hand sides of the sum rules \eqref{spinSR} in their valid Borel regions. The bands show the errors as obtained from the Monte Carlo--based analysis. 

\begin{table}[t]
	\caption{The isoscalar ($g_A^q$ and $g_A^s$) axial-vector coupling constants of $\Lambda$ as obtained from QCDSR. For comparison, we also give the coupling constants from naive SU(3)$_F$ assuming $\langle N | \bar{s}\gamma_\mu \gamma_5 s |N\rangle=0$ (denoted by SU(3)$_F$[naive]), and those from DIS data assuming SU(3)$_F$ (denoted by SU(3)$_F$[DIS]), which are obtained by inserting $\Sigma=3F-D$ and $\Sigma=0.33$ (central value as reported by the HERMES Collaboration~\cite{Airapetian:2007mh}), respectively, into the SU(3)$_F$ relations. As for the SU(3)$_F$ parameters we use $F/D=0.575$ and $F+D=1.269$~\cite{Amsler:2008zzb}.}
\begin{center}
\begin{tabular*}{0.45\textwidth}{@{\extracolsep{\fill}}lccccccc}
		\hline\hline 
		$g_A$  & SU(3)$_F$[naive] & SU(3)$_F$[DIS] & QCDSR \\[0.5ex]
		\hline 
		$g_A^q$ &  $-0.15$ & $-0.32$ & $-0.37\pm 0.13$ \\
		$g_A^s$ & 0.73 & $0.65$ & $0.51 \pm 0.11$\\
		\hline\hline
\end{tabular*}
	\label{su3qc}
\end{center}
\end{table}

Our numerical results are given in Table~\ref{su3qc}. For comparison, we also give the coupling constants from SU(3)$_F$ assuming $\langle N | \bar{s}\gamma_\mu \gamma_5 s |N\rangle=0$, and those from DIS data assuming SU(3)$_F$ relations given as $g_A^q=2/3(\Sigma-D)$ and $g_A^s=1/3(\Sigma+2 D)$, where $\Sigma$ is equivalent to the flavor-singlet axial-vector coupling constant, $g^0_A$. The value we obtain for $g_A^q$ in QCDSR, namely $g_A^q=-0.37 \pm 0.13$, is in nice agreement with the experimental result, while $g_A^s$ lies slightly lower. 

\begin{figure}
	[th] 
	\caption{Same as Fig.~\ref{axials} but for $g_A^q$ in the SU(3)$_F$-broken case.}
	\includegraphics[scale=0.50]{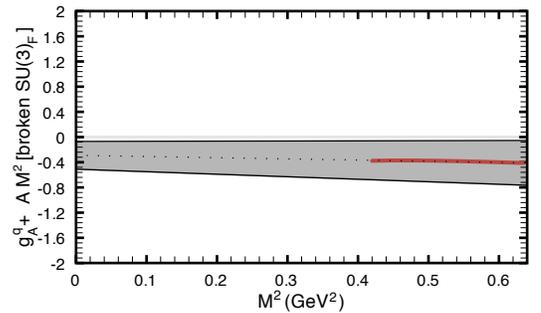} \label{axials_su3b}
\end{figure}

The SU(3)$_F$-breaking effects are accounted for by restoring the physical values of the parameters $m_s$ and $f$ in the sum rules \eqref{spinSR} and \eqref{massSR}. We consider only the sum rule for $g_A^q$ since the susceptibilities associated with this coupling are unaffected with SU(3)$_F$ breaking. We apply a similar procedure as above where we obtain $\tilde{\lambda}_\Lambda$ from the mass sum rule in~\eqref{massSR} with the SU(3)$_F$-breaking effects and this value is used in the sum rules of $g_A$ for each corresponding parameter set. In Fig.~\ref{axials_su3b}, we plot the left-hand and the fitted right-hand side of the sum rule~\eqref{spinSR} for $g^q_A$ in the SU(3)$_F$-broken case. We obtain $g_A^q=-0.29 \pm 0.22$, which is consistent with the value obtained in the SU(3)$_F$ limit. This implies that the SU(3)$_F$-breaking effects are small and the contributions from the $u$ and the $d$ quarks to the $\Lambda$ polarization are negatively polarized as in the SU(3)$_F$ limit. 

\section{Conclusion}\label{conc}
In conclusion, we have calculated the isoscalar axial-vector coupling constants of $\Lambda$ using the method of QCDSR. This information reveals the individual contributions of the $u$, $d$ and $s$ quarks to the spin content of $\Lambda$. We have found that in the SU(3)$_F$ limit our results for $g_A^q$ are in agreement with expectations based on experiment assuming SU(3)$_F$ symmetry while the value we obtain for $g_A^s$ slightly deviates from the empirical one. We have also analyzed the isoscalar coupling $g_A^q$ with SU(3)$_F$ breaking effects and have found that the light-quark contributions remain mainly unaffected and negatively polarized as in the SU(3)$_F$ limit.

\appendix
\section{}\label{appA}
A simple model for the $\Lambda$ baryon is that it is contaminated by  $\Sigma$ baryon together with a $\pi$ having orbital angular momentum $L=1$ in order to conserve parity. In the naive quark model the $\Sigma$ is composed of a $u$-$d$ quark pair coupled to $S=1$ and a $s$-quark with $S=1/2$. In this picture, the quark-spin configurations of $\Sigma$ are given as
\begin{align}
	\begin{split}\label{spconf1}
		|\Sigma(+1/2)\rangle &= \sqrt{2/3}|[ud](+1),s(-1/2)\rangle \\
		&\quad- \sqrt{1/3}|[ud](0),s(+1/2)\rangle,	
	\end{split}\\
	\begin{split}\label{spconf2}
		|\Sigma(-1/2)\rangle &= -\sqrt{2/3}|[ud](-1),s(+1/2)\rangle \\
		&\quad+ \sqrt{1/3}|[ud](0),s(-1/2)\rangle,
	\end{split}
\end{align}
using the appropriate Clebsch-Gordan coefficients. Similarly, the $\Sigma$-$\pi$ mixed state is written as
\begin{align}
	\begin{split}\label{mixst}
		|\Sigma,\,\pi;\,J=1/2\rangle = -\sqrt{2/3}|\Sigma(-1/2), \,L_\pi(+1)\rangle \\
		\quad+\sqrt{1/3}|\Sigma(+1/2),\,L_\pi(0)\rangle.
	\end{split}
\end{align}
Inserting the spin configurations in Eqs.~\eqref{spconf1} and~\eqref{spconf2} into Eq.~\eqref{mixst}, we obtain
\begin{align}
	\begin{split}
		&|\Sigma,\,\pi;\,J=1/2\rangle =\\
		&\quad 2/3|[ud](-1),\, s(+1/2),\,L_\pi(+1)\rangle\\
		&\quad -\sqrt{2}/3 |[ud](0),\,s(-1/2),\,L_\pi(+1)\rangle\\
		&\quad +\sqrt{2}/3 |[ud](+1),\,s(-1/2),\,L_\pi(0)\rangle\\
		&\quad -1/3 |[ud](0),\,s(+1/2),\,L_\pi(0)\rangle.
	\end{split}
\end{align}
It is then straightforward to calculate the spin probabilities of the $u$-$d$ quark pair in the $\Sigma$-$\pi$ mixed state as $2/9$, $3/9$ and $4/9$ corresponding to $[ud](+1)$, $[ud](0)$ and $[ud](-1)$ configurations, respectively, which results in an expectation value of $-2/9$ for the spin of the $u$-$d$ pair.
\acknowledgments
This work was supported in part by KAKENHI (17070002 and 19540275).


\end{document}